\documentclass[prl,aps,reprint,footinbib,longbibliography]{revtex4-1}
\usepackage{graphicx}
\usepackage{amssymb}
\usepackage{amsmath}
\usepackage{times}

\begin{document}

\title{Enlarged Kuramoto Model: Secondary 
Instability and Transition to Collective Chaos}

\author{Iv\'an Le\'on}
\author{Diego Paz\'o}
\affiliation{Instituto de F\'{\i}sica de Cantabria (IFCA), Universidad de 
Cantabria-CSIC, 39005 Santander, Spain}

\date{\today}

\begin{abstract}
The emergence of collective synchrony from an incoherent state
is a phenomenon essentially described by the Kuramoto model.
This canonical model was derived perturbatively, by applying 
phase reduction to an ensemble of heterogeneous, 
globally coupled Stuart-Landau oscillators. 
This derivation neglects nonlinearities in the 
coupling constant.
We show here 
that a comprehensive analysis requires extending
the Kuramoto model up to quadratic order. This 
`enlarged Kuramoto model' comprises three-body (nonpairwise) interactions, 
which induce strikingly complex phenomenology at certain parameter values.
As the coupling is increased, a secondary instability renders the synchronized state unstable, 
and subsequent bifurcations lead to collective chaos.
An efficient numerical 
study of the thermodynamic limit, valid for Gaussian heterogeneity, is carried out by means of a Fourier-Hermite decomposition
of the oscillator density.
\end{abstract}

  \maketitle

Collective synchronization is a phenomenon in which an ensemble
of heterogeneous, self-sustained oscillatory units (commonly known as oscillators) spontaneously
entrain their rhythms. 
This is a pervasive phenomenon observed in natural systems and man-made devices, 
covering a wide range of spatio-temporal scales, from
cell aggregates to swarms of fireflies \cite{Str03,PRK01}.
  
Seeking to understand the onset of collective synchronization, Winfree 
invented a model
consisting of globally coupled 
oscillatory units with one degree of freedom
(phase oscillators) \cite{Win67,Win80}. 
Following this scheme, Kuramoto found an analytically tractable model, which
captures the onset of collective synchronization from an incoherent
state \cite{Kur75,Kur84}. 
Due to its simplicity, 
the Kuramoto model and its generalization with phase-lagged coupling
---the so-called Kuramoto-Sakaguchi model after Ref.~\cite{SK86}---,
have been intensely studied,
with a vast number of extensions 
and applications in several fields \cite{ABP+05,rodrigues16}.

The Kuramoto(-Sakaguchi) model is often introduced as above, 
i.e.~as a mere mathematical refinement of the Winfree model.
However, this is only partly true, since 
Kuramoto rigorously derived the model bearing his name. In particular,
he applied phase reduction to an ensemble of weakly
coupled Stuart-Landau oscillators \cite{Kur75,Kur84}. 
The Stuart-Landau oscillator is a relevant natural choice,
as it represents a generic limit-cycle attractor close to a Hopf bifurcation. 

Kuramoto's perturbative phase-reduction approach 
is valid for weak coupling. Specifically, oscillator heterogeneity
and interactions appear at zeroth and linear orders in the coupling constant, respectively.
These considerations explain why the 
quadratic order was neglected in the original Kuramoto model.
Nevertheless, in certain circumstances, 
going beyond the first (or linear) order may be required.
Indeed, the description of some experiments with lattices of 
optomechanical \cite{marquadt11} and
nanoelectromechanical \cite{matheny19} oscillators rely on second-order
phase reductions. The analysis of the corresponding
second-order phase-reduced models has
remained, however, rather incomplete. The reason for this is the
nonpairwise interactions appearing at quadratic order. 
From this perspective, the original setup with heterogeneous, 
diffusively coupled 
Stuart-Landau oscillators 
appears to be the ideal testbed model for investigating
second-order phase reduction to the fullest extent possible.
So far, only the  case of identical oscillators 
has been analyzed \cite{leon19}.

Recently, nonpairwise (also called `higher-order') interactions  
are attracting growing attention 
in several fields, such as neuroscience, ecology, and social systems
(see Refs.~\cite{stankovsky_rmp17,battiston20} and references therein). In this spirit, 
several works 
have considered populations
of phase oscillators with nonpairwise interactions from the outset.
Simplifying ad-hoc assumptions, such as 
absent pairwise coupling \cite{tanaka11,KP15,xu21,wang21}
and/or particularly convenient nonpairwise interactions
\cite{SA19,SA20,lucas20,wang21} (e.g.~admitting the Ott-Antonsen ansatz \cite{OA08}), are 
adopted seeking analytical tractability. 

In this Letter we extend the Kuramoto model up to second order 
in the coupling constant $\epsilon$. In this ``enlarged'' Kuramoto model 
the new terms of order $\epsilon^2$ comprise two different three-body (nonpairwise) interactions.
Strikingly, their combined 
action triggers a secondary instability in which
standard collective synchronization destabilizes.
This is the precursor of a sequence of instabilities giving rise to 
a state of collective chaos.
We efficiently investigate the thermodynamic limit of the model by
means of a Fourier-Hermite decomposition of the oscillator density. 
This scheme appeared 
some
years ago in a theoretical study ~\cite{chiba13}, 
but it is numerically implemented here for the first time (adopting an appropriate closure).

The starting point of our work
is a heterogeneous population of $N\gg1$ Stuart-Landau oscillators with global diffusive coupling: 
 \begin{equation}\label{CGLE}
 \dot{A}_j= (1+i\sigma\omega_j)A_j - ( 1 + i c_2) |A_j|^2 A_j+\epsilon(1+ic_1) \left(\overline{A}- A_j\right). 
 \end{equation}
Here $A_j\equiv r_j e^{i\phi_j}$ is a complex variable, and index $j$ runs from 1 to $N$.
The $\omega_j$'s are drawn from a unit-variance normal distribution 
$g(\omega)$. 
The mean of $g(\omega)$ is selected to be 0, by going to a rotating frame if necessary.
Therefore, each individual Stuart-Landau oscillator possesses a natural 
frequency equal to $\sigma\omega_j-c_2$,
where $c_2$ is the noniscochronicity parameter. 
Parameter $\sigma>0$ is included to account for the frequency dispersion.
Concerning the coupling, it is diffusive through the mean field 
$\overline{A}=\frac{1}{N}\sum_{i=1}^{N}A_i$.
Parameter $\epsilon>0$ controls the coupling strength,
and $c_1$ modulates its reactivity. We are exclusively interested in the thermodynamic 
limit ($N\to\infty$) of the model.
In this work we select
$\sigma=10^{-3}$
and $c_2=3$ (a standard value in the literature, 
see e.g.~\cite{NK93,*CP19}), leaving $c_1$ and $\epsilon$ as control parameters. 
The effect of varying $c_2$ and $\sigma$ is  discussed at the
end of this Letter.

\begin{figure}
	\includegraphics[width=\linewidth]{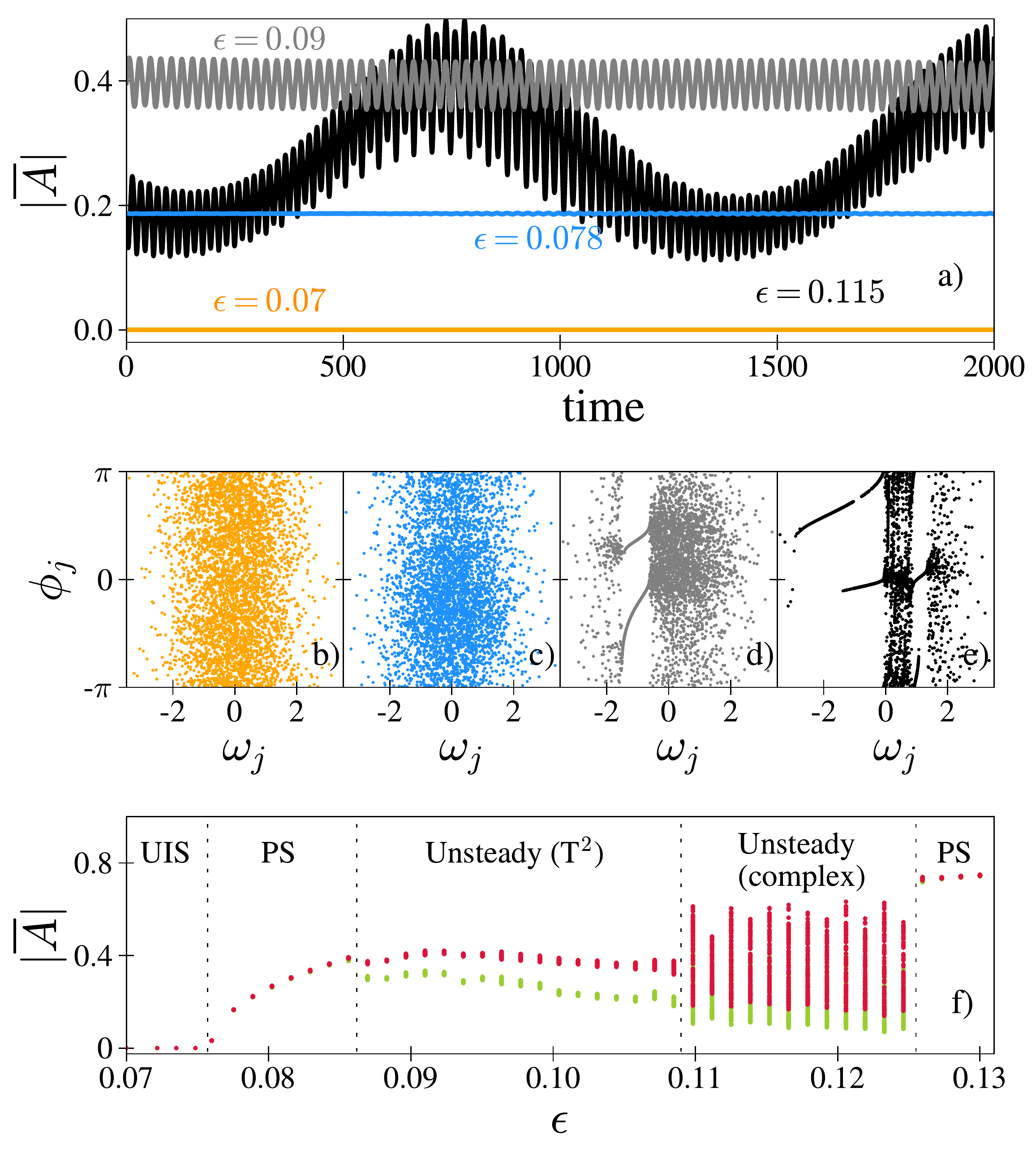}
	\caption
{Dynamics of the population of 20000 Stuart-Landau oscillators, Eq.~\eqref{CGLE},
for different values of $\epsilon$ with $c_1=-0.4$, $c_2=3$, and $\sigma=10^{-3}$.
(a) Time series of the mean field amplitude 
$|\overline A|$ for $\epsilon=0.07$, $0.078$, $0.09$, and $0.115$. 
$|\overline A|\simeq0$, 
$|\overline A|\simeq\mathrm{const.}>0$, and periodic $|\overline A(t)|$
correspond to UIS, PS, and quasiperiodic global attractor, respectively.
(b,c,d,e) Snapshots of the angular variables $\phi_j$ for each of the
four $\epsilon$ values chosen in (a). Only a subset of 4000 oscillators are shown for clarity. 
(f) Local maxima and minima of $|\overline{A}|$
as constant $\epsilon$ is increased by steps of size $1.35\times10^{-3}$.}
\label{figestado}
\end{figure}

\begin{figure*}
	\includegraphics[width=\linewidth]{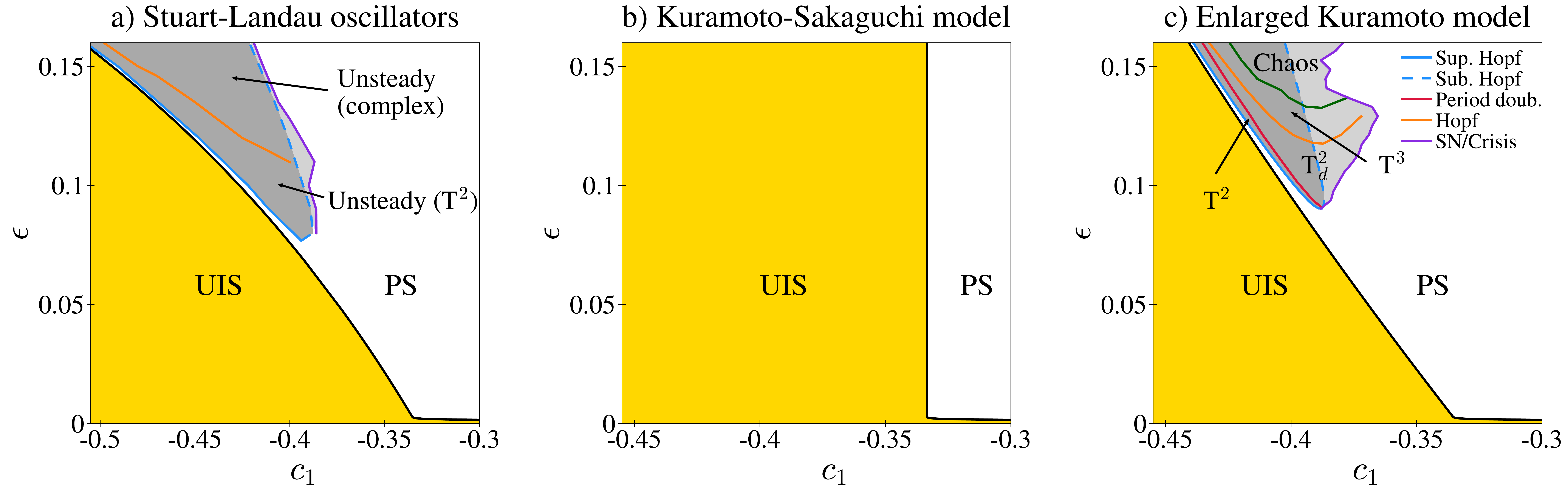}
	\caption{Phase diagrams of model (1) for $c_2=3$ and $\sigma=10^{-3}$, 
	as well as
	its first- and second-order phase
	reductions.
	(a) Model \eqref{CGLE}: all boundaries were obtained 
	from numerical simulations with a population of $N=20000$ Stuart-Landau oscillators, 
	save the boundary of UIS (obtained analytically). In the dark shaded region UIS and PS are both unstable, and
	$|\overline A|$ varies with time. 
	In the light shaded region PS coexists with an unsteady state.
	(b) Kuramoto-Sakaguchi model obtained from Eq.~\eqref{eqsecondord} discarding quadratic terms in $\epsilon$.
	(c) Enlarged Kuramoto model, Eq.~\eqref{eqsecondord}; all boundaries, except the UIS-PS line, were determined
	using Eq.~\eqref{odes}. The right boundary of the bistability region (in purple color) indicates 
	where the attractor with unsteady dynamics abruptly disappears, indistinctively through
	a saddle-node bifurcation of tori,
	a boundary crisis, or any other bifurcation.}
	\label{figdiagram}
\end{figure*}

System \eqref{CGLE} displays a plethora of complex states. In particular,
collective chaos already emerges at moderate and large coupling 
under simplifying assumptions
such as, homogeneity ($\sigma=0$) ~\cite{HR92,NK93,*CP19}
and vanishing reactivity and shear ($c_1=c_2=0$) \cite{MS90,*MMS91}. 
We focus here on the weak coupling regime, in which 
the oscillators remain close to their original limit cycles
at $r_j=1$ and a phase description becomes possible. 
Two states are generically 
expected for small $\epsilon$. 
On the one hand, there is the uniform incoherent state (UIS),
corresponding to a vanishing 
mean field $\overline A$ (in the thermodynamic limit), with 
the oscillators angles $\phi_j$ uniformly scattered, see Figs.~\ref{figestado}(a)
and \ref{figestado}(b) for particular 
parameter values and $\epsilon=0.07$. 
On the other hand, typically, as
$\epsilon$ exceeds a certain threshold UIS becomes unstable and a
state of collective partial synchrony (PS) emerges. In this configuration, a macroscopic proportion
of the oscillators becomes entrained to a common frequency
$\langle{\dot\phi}_{j\in S}\rangle=\Omega$
and the mean field rotates uniformly with constant amplitude: $|\overline A|=\mathrm{const}$. In a finite population, as in Fig.~\ref{figestado}(c),
entrained oscillators may not be observed, since they belong to one of the tails
of $g(\omega)$. Drifting oscillators alone
cause $\overline A$ to depart from zero. 
Surprisingly, our numerical simulations
indicate that the dynamics may become
of a different kind as the coupling is further 
 increased, while still remaining small. 
As shown in Fig.~\ref{figestado}(a) for $\epsilon=0.09$, the
 collective dynamics incorporates a new frequency, 
and $|\overline{A}(t)|$ oscillates periodically, i.e.~the 
attractor is a two-dimensional
torus or $\mathrm{T}^2$ (disregarding finite-size fluctuations).
Figure~\ref{figestado}(d) shows the corresponding snapshot of the angles $\phi_j$ for $\epsilon=0.09$.
We may see that part of the population forms a two-cluster state that evolves in time
such that the phase differences are time-dependent but bounded. 
As far as we know, this unsteady configuration with time-dependent clusters  
has not been observed before in Eq.~\eqref{CGLE}. 
It is very much alike the Bellerophon state coined 
 in ~\cite{bellerophon,*bellerophon2} for ensembles of phase oscillators. 
For still larger $\epsilon$,
$|\overline A|$ exhibits even more complex oscillations,
as can be seen setting $\epsilon=0.115$ in Fig.~\ref{figestado}(a). 
In Fig.~1(f) we represent the local maxima
and minima of $|\overline A(t) |$ as a function of $\epsilon$. 
The low-frequency modulation sets in at $\epsilon\approx0.109$.
As a result of the instability, a three-frequency quasiperiodic 
collective motion is, in principle, expected. Still, an additional transition to
weak collective chaos cannot be ruled out.
At some parameter values (e.g.~$\epsilon=0.14$, $c_1=-0.415$), see the Supplemental Material, 
the largest Lyapunov exponent does not decay to zero with the system size,
what is a clear indication of collective chaos. (For the value $\epsilon=0.115$ taken in 
Fig.~1 the result is inconclusive.)

To put the previous observations in a wider framework we numerically
determined where the unsteady behavior occurs in the $c_1-\epsilon$ plane.
The phase diagram in Fig.~\ref{figdiagram}(a)
shows where qualitatively different dynamics are observed.
The stability boundary of UIS was analytically computed 
following the approach in \cite{cross06},
see the Supplemental Material. Remarkably, numerical simulations
of Eq.~\eqref{CGLE} reveal that 
PS is unstable
inside the dark shaded region in Fig.~\ref{figdiagram}(a), i.e.~unsteady $|\overline A(t)|$ 
spontaneously sets in. In addition, numerical continuation discloses 
an adjacent narrow band of coexistence between unsteady dynamics and PS. 
The orange line in Fig.~2(a) divides the unsteady region into two parts: the lower one 
with $\mathrm{T}^2$ collective motion, and the upper one with more complex oscillations. 
We emphasize that determining the exact nature of the complex unsteady states 
is an arduous work, which hinders a more detailed phase diagram.

At this point, we resort to phase reduction in order to 
 better understand the nature and organization of the unsteady collective states.
For weak coupling phase reduction
allows us to describe the system 
solely in terms of phase variables $\theta_j=\phi_j-c_2 \ln r_j$ \cite{PRK01,Kur84}.
Following \cite{leon19} we write down the 
second-order phase reduction \footnote{A complete derivation can be found in the Supplemental Material. In Eq.~\eqref{eqsecondord} terms of orders $O(\epsilon^3)$ and $O(\epsilon^2\sigma)$ have been neglected, since small coupling strength ($\epsilon\ll1$) and frequency dispersion ($\sigma\ll1$) are assumed. 
Moreover, note that phase reduction does not introduce terms proportional to $\epsilon\sigma$, as in \cite{gengel21}.}
of \eqref{CGLE}, or `enlarged
Kuramoto model':
 \begin{widetext}
 \begin{equation} \label{eqsecondord}
 \dot{\theta}_j=\sigma\omega_j+
 \epsilon \eta \,  R  \, \sin(\Psi-\theta_j+\alpha) 
 + \frac{\epsilon^2 \eta^2}{4} 
 \left[R\sin(\Psi-\theta_j+\beta)
 - R^2 \sin(2\Psi-2\theta_j+\beta)+   R \, Q \sin(\Phi-\Psi-\theta_j) 
  \right] ,
 \end{equation}  
\end{widetext} 
where three new constants, depending on $c_1$ and $c_2$, are defined: 
$\eta\equiv\sqrt{(1+c_2^2)(1+c_1^2)}$; and the phase lags 
$\alpha\equiv\arg[1+c_1 c_2+(c_1-c_2)i]$, and
$\beta\equiv\arg(1-c_1^2+2 c_1 i)$. For simplicity,
we have chosen a reference frame with vanishing central frequency.
Interactions involve two mean fields, $Z_1\equiv R \, e^{i\Psi}$ and $Z_2\equiv Q \, e^{i\Phi}$, 
which are 
the first two elements of  an infinite set of  Kuramoto-Daido order
parameters \cite{Dai93}: 
$Z_k\equiv N^{-1} \sum_{j=1}^N e^{i k \theta_j}$.
  Equation \eqref{eqsecondord} includes nonpairwise interactions, 
which are inherent to higher-order phase reduction,
even if the coupling in the original system \eqref{CGLE} is pairwise and linear 
\cite{leon19,KuraNakao19,gengel21}.
 In particular, three-body interactions are conveyed by
the last two terms 
\footnote{Notice these identities:
$R^2 \sin(2\Psi-2\theta_j+\beta)=N^{-2}
\sum_{k,l} \sin(\theta_k+\theta_l-2\theta_j+\beta)$ ; 
$R \, Q \sin(\Phi-\Psi-\theta_j)=N^{-2}   
\sum_{k,l}  \sin(2\theta_k-\theta_l-\theta_j)$}
and are comparatively weak (of order $\epsilon^2$),
as usual in physics \cite{hammer13}. 
This is not the case of 
most previous studies on coupled phase oscillators \cite{tanaka11,KP15,SA19,SA20,xu21,
bick16,SA20_memory}, but see \cite{matheny19,leon19,gengel21,rosenblum21}.

We start the analysis of Eq.~\eqref{eqsecondord} noticing that if we neglect the $O(\epsilon^2)$ terms, then 
we recover the Kuramoto-Sakaguchi model with coupling constant $\epsilon \eta $.
For $N\to\infty$, the phase diagram resulting from this 
$O(\epsilon)$ approximation
is shown in Fig.~\ref{figdiagram}(b). 
The only attracting configurations 
are UIS and PS. 
The boundary of UIS can be calculated following \cite{SK86}. 
It diverges at $c_1=-c_2^{-1}=-1/3$, corresponding to $\alpha=-\pi/2$.
When comparing Figs.~\ref{figdiagram}(a) and \ref{figdiagram}(b), it is 
manifest that  first-order phase reduction does not
provide a faithful description of system \eqref{CGLE}
in the left part of the phase diagram.

We now consider Eq.~\eqref{eqsecondord} in full. 
Concerning the linear stability of UIS ($R=Q=0$),
only the first term of order $\epsilon^2$ is relevant. 
It may be added to the linear term to recalculate the stability boundary \cite{SK86}, see the Supplemental Material. 
The result is shown as a solid black line in Fig.~\ref{figdiagram}(c). Now the 
boundary of UIS exhibits a knee at $c_1\approx-1/3$, in qualitative agreement with Fig.~\ref{figdiagram}(a). Analyzing the stability of PS is a much harder problem.
Through a numerical self-consistent approach \cite{SK86} we tracked
the branch of PS emanating from incoherence. However, 
this does not allow us to determine
its stability.
Moreover, the direct numerical integration of Eq.~\eqref{eqsecondord}
is not  more efficient than simulating 
Eq.~\eqref{CGLE}: The number
of degrees of freedom is reduced by a factor 2,
but at the cost of including computationally expensive trigonometric functions.

  In order to exploit the 
dimensionality reduction achieved in Eq.~\eqref{eqsecondord},
an alternative strategy is required.
We resort to a moments 
system introduced almost a decade
ago by Chiba in his 
theoretical study of the Kuramoto model \cite{chiba13}.
Crucially, working with a set of moments avoids 
finite-size fluctuations and the concomitant microscopic (phase) chaos \cite{popovych05}.
We start defining the density 
$\rho(\theta|\omega,t)$, such that 
$\rho(\theta|\omega,t)d\theta$ is the fraction of oscillators with phases 
between $\theta$ and $\theta+d\theta$ and frequency $\omega$ at time $t$.
Now, we write the Fourier-Hermite decomposition of $\rho$:
\begin{equation}
 \rho(\theta|\omega,t)=
 \frac1{2\pi} \sum_{k=-\infty}^\infty \sum_{m=0}^\infty 
 P_k^m(t) e^{-ik\theta} h_m(\omega) ,
 \label{expansion}
\end{equation}
where $h_m(x)=\mathrm{He}_m(x)/\sqrt{m!}$ are normalized (probabilist's) Hermite polynomials:
$\int_{-\infty}^\infty h_m(\omega) h_n(\omega) g(\omega) d\omega=\delta_{mn}$. 
The Fourier-Hermite coefficients $P_k^m$ are  obtained inverting 
Eq.~\eqref{expansion}:
	\begin{equation}\label{defmod}
	P^m_k(t)=
	\int_0^{2\pi} d\theta e^{ik\theta} \int_{-\infty}^\infty  d\omega 
	h_m(\omega) g(\omega)\rho(\theta|\omega,t)  .
	\end{equation}
These Fourier-Hermite modes extend the 
Kuramoto-Daido order parameters to the space of the
natural frequencies. Specifically, $P_k^0=Z_k$ (in the $N\to\infty$ limit). 
The density $\rho$ obeys the continuity equation $\partial_t\rho=-\partial_{\theta}(\rho\, \dot{\theta})$. Inserting the expansion \eqref{expansion},
using the recurrence relation
$\omega h_m=\sqrt{m} h_{m-1}+\sqrt{m+1} h_{m+1}$ \cite{AS72},
and redefining 
$P_k^m\to (-i)^m P_k^m$ for convenience, we get
an infinite set of ordinary differential equations:
	\begin{eqnarray}
    {k}^{-1}\dot{P}_k^m&=& \sigma\left(\sqrt{m} P_k^{m-1}-\sqrt{m+1}P_k^{m+1}\right) \nonumber\\
	&+&\frac{\epsilon \eta}{2}\left(P_{k-1}^mZ_1e^{i\alpha}-P_{k+1}^m Z_1^* e^{-i\alpha} \right) \nonumber\\
    &+&\frac{\epsilon^2\eta^2}{8}\left(
    P_{k-1}^m Z_1 e^{i\beta}-P_{k+1}^m Z_1^* e^{-i\beta}-
		P_{k-2}^m Z_1^2 e^{i\beta} \right.\nonumber \\ 
		&+& \left.P_{k+2}^m Z_1^{*2} e^{-i\beta}+
	P_{k-1}^m Z_2 Z_1^* -P_{k+1}^m Z_2^* Z_1 \right),
	\label{odes}
	\end{eqnarray}
where the asterisk denotes complex conjugation. 
System \eqref{odes} is 
equivalent to Eq.~\eqref{eqsecondord} with $N\to\infty$.

The numerical integration of Eq.~\eqref{odes} requires 
to implement a truncation at finite $k_\mathrm{max}$
and $m_\mathrm{max}$, with an adequate closure.
Note first that, in the UIS, $P_0^0=1$ is the only nonzero coefficient,
whereas in the PS state the modes
decay with $k$ and $m$ roughly as $|P_k^m|\sim e^{-a k} e^{-b\sqrt{m}}$.
We imposed the boundary conditions:
$P_{k_\mathrm{max}+1}^m=0$, and $P_k^{m_\mathrm{max}+1}=2P_k^{m_\mathrm{max}}-P_k^{m_\mathrm{max}-1}$.
We tested the performance of different system sizes, 
finding that $k_\mathrm{max}=m_\mathrm{max}=40$
already yield an excellent convergence, even for strongly unsteady states. 
Therefore, our analysis 
below relies on Eq.~\eqref{odes} with $n_f= k_\mathrm{max}
\times (m_\mathrm{max}+1)\times2=3280$ degrees of freedom. 
In comparison, simulating Eq.~\eqref{eqsecondord} with $n_f$ oscillators
is unproductive because of unavoidable finite-size fluctuations.

One now can see that the PS state corresponds to a solid rotation 
$P_k^m(t)= p_k^m e^{ik\Omega t}$. After inserting 
this solution into Eq.~\eqref{odes}, the unknowns $p_k^m$ and
$\Omega$ are found
via a Newton-Raphson algorithm (imposing $p_1^1\in\mathbb R$). 
The result completely agrees with the one obtained from 
the self-consistent numerical calculation mentioned above. 
Now, however, we can determine linear stability.
Moving to a rotating frame with angular 
velocity $\Omega$, we 
linearize the system around the fixed point. The locus of a
secondary (Hopf) instability
is accurately located requiring the eigenvalues of the Jacobian matrix
with the largest real part 
to be $\pm i\Omega_H$ (with an extra zero eigenvalue due to 
rotational invariance $P_k^m\to e^{ik\gamma} P_k^m$).
The Hopf line is shown 
in blue in Fig.~\ref{figdiagram}(c).
The transition is supercritical (subcritical) at
the solid (dashed) line.
The emerging oscillatory mode yields a torus attractor ($\mathrm{T}^2$), 
in which, due to the rotational symmetry, 
no lockings on its surface are expected, see e.g.~\cite{rand82,barkley94}.
Recalling Eq.~\eqref{eqsecondord} we infer that,
at the level of the individual oscillators, 
the superimposed oscillation induces entrainment 
at frequencies $\Omega+ (n/2) \Omega_H$ 
($n\in\mathbb{Z}$).
The half-integer frequency plateaus stem from the 
term accompanying $R^2$ in Eq.~\eqref{eqsecondord}.
In particular, 
the two clusters in Fig.~1(d) 
correspond to a frequency plateau at frequency $\Omega+ \Omega_H/2$.

The remaining regions of the phase diagram in Fig.~\ref{figdiagram}(c)
are determined from 
direct numerical simulations of Eq.~\eqref{odes} with the
aforementioned closure, as well as 
by computing the largest Lyapunov exponents $\{\lambda_i\}_{i=1,2,\ldots}$. 
Our systematic exploration reveals a period-doubling bifurcation line 
($\mathrm{T}^2\to \mathrm{T}_d^2$ transition)
close to the supercritical-Hopf line.
The period-doubling bifurcation line almost certainly exists also for 
the ensemble of Stuart-Landau oscillators.
Magnifying the gray line in Fig.~1(a) the signature of a doubled torus $\mathrm{T}_d^2$
can be discerned. However, it is very hard to determine the bifurcation point due to the long transients involved 
and unavoidable finite-size fluctuations, see Fig.~1(f). 

As occurs with the ensemble of Stuart-Landau oscillators, the torus attractor undergoes a
Hopf bifurcation, see the orange line in Fig.~2(c). 
Thereby three-frequency quasiperiodic dynamics
($\mathrm{T}^3$ attractor) emerges, consistently with three vanishing Lyapunov exponents.


Adjacent to the $\mathrm{T}^3$ domain in Fig.~2(c), there exists a region with
chaotic dynamics, in conformity with the Ruelle-Takens-Newhouse scenario.
As occurred with system \eqref{CGLE}, Fig.~\ref{figdiagram}(a), 
PS and unsteady states coexist. In Fig.~\ref{figdiagram}(c) the bistability region
is bounded
by a purple line denoting either a saddle-node 
bifurcation, emanating from a (codimension-2) Bautin point
at the bottom of the Hopf line, or an attractor crisis.
The phase diagram in Fig.~\ref{figdiagram}(c)
 reveals  which are the unsteady collective states of \eqref{CGLE}, 
and their expected arrangement.
Indeed, obtaining 
a phase diagram with the degree of detail of Fig.~\ref{figdiagram}(c) 
is virtually unattainable simulating
the original system, Eq.~\eqref{CGLE}.

\begin{figure}
	\includegraphics[width=\linewidth]{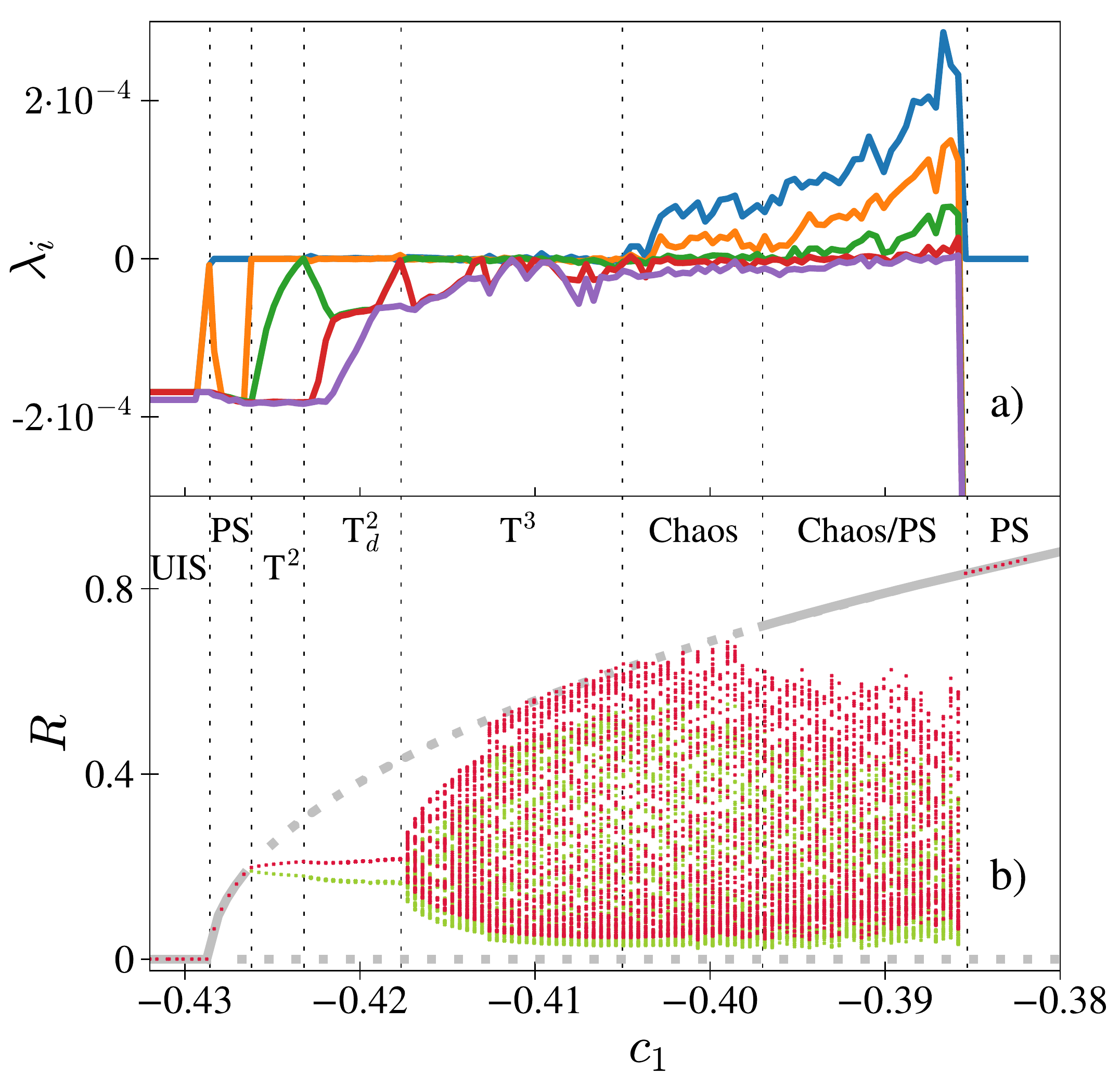}
	\caption{Sequence of bifurcations of Eq.~\eqref{eqsecondord}, 
	 obtained from Eq.~\eqref{odes},
	as $c_1$ is increased with $\epsilon=0.14$. 
	(a) Five largest Lyapunov exponents $\{\lambda_i\}_{i=1,\ldots,5}$.
	(b) Local maxima and minima of $R(t)$. 
	As a reference, the $R$ values of  
	UIS ($R=0$) and PS ($R>0$) are depicted in gray. Solid (dashed)
	lines correspond to linearly stable (unstable) states.
}
	\label{figbif}
\end{figure}

To better characterize the chaotic region, a detailed exploration along the 
horizontal line $\epsilon=0.14$ is shown in Fig.~\ref{figbif}.
In Figs.~\ref{figbif}(a) and ~\ref{figbif}(b) the five largest Lyapunov
exponents and the local maxima and minima of $|P_1^0(t)|=R(t)$
are, respectively, depicted for the same $c_1$ range. In the $\mathrm{T}^3$ 
interval there may be some additional bifurcations (lockings or torus doubling),
which we did not attempt to resolve. Interestingly, in the chaotic domain
an increasing number of Lyapunov exponents become positive
as $c_1$ increases, i.e.~collective chaos transforms into 
collective hyperchaos.

In this Letter we have introduced the `enlarged Kuramoto model';
a population of phase oscillators in which three-body interactions 
enter in a perturbative way. Remarkably, this 
makes a world of difference, 
drastically reshaping the traditional Kuramoto scenario.
The `enlarged Kuramoto model'
exhibits a variety of unsteady states, including
collective chaos and hyperchaos. To our knowledge,
these states
have not been previously reported in a population
of   globally coupled phase oscillators, with 
a unimodal distribution of the natural frequencies.
We have considered a particular frequency dispersion $\sigma=10^{-3}$
in Fig.~\ref{figdiagram}(c). 
If $\sigma$ is lowered the bottom of the Hopf bifurcation 
line approaches the $c_1$ axis at $c_1=-c_2^{-1}$. 
This is expected to occur for any nonzero $c_2$ value,
in consistence with  the 
$\sigma=0$ case \cite{leon19} (to be shown elsewhere). 
Nonetheless, only heterogeneity, in contradistintion to weak noise \cite{leon19,leon20},
is able to trigger
unsteady collective dynamics (absent for $\sigma=0$).
As a final remark, we stress that 
reducing the population of
Stuart-Landau oscillators \eqref{CGLE} to
the phase model \eqref{eqsecondord} is both illuminating and convenient, 
as it enables an efficient investigation of the
thermodynamic limit by virtue of the Fourier-Hermite expansion.
The application of this scheme to other populations
of 
phase oscillators with Gaussian heterogeneity is straightforward. 
For other forms of $g(\omega)$ the suitable set of orthogonal
polynomials must be adopted: e.g.~the Fourier-Legendre mode
decomposition is appropriate for uniform $g(\omega)$.

\begin{acknowledgments}

 We thank Ernest Montrbri\'o and Juan M.~L\'opez for their
critical reading of the manuscript.
We acknowledge support by Agencia Estatal de Investigaci\'on and 
Fondo Europeo de Desarrollo Regional under Project 
No.~FIS2016-74957-P (AEI/FEDER, EU).
IL acknowledges support by Universidad de Cantabria and 
Government of Cantabria
under the Concepci\'on Arenal programme.
\end{acknowledgments}

\begin{figure*}
	\includegraphics[page=1,width=\textwidth]{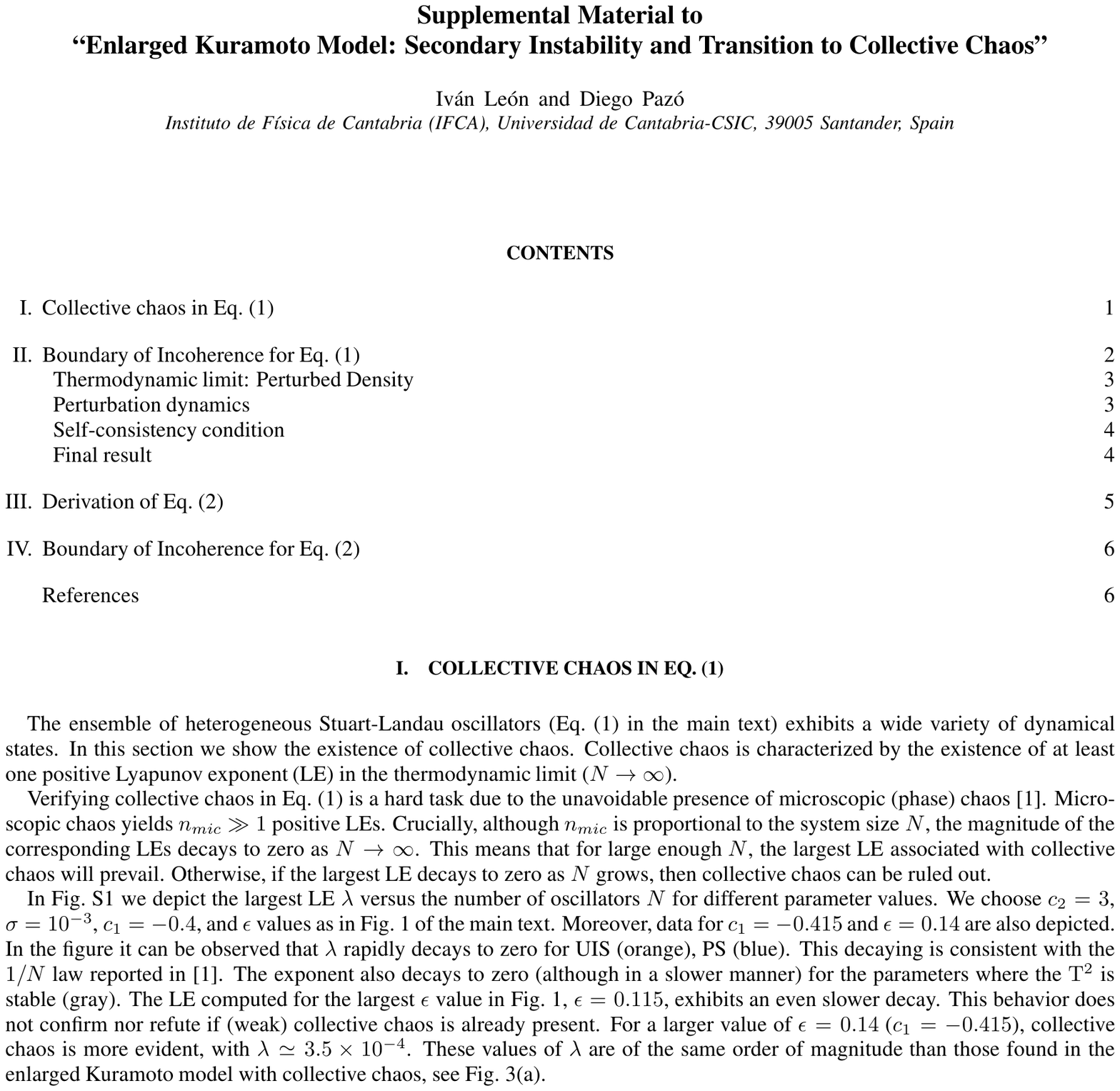}
\end{figure*}
\begin{figure*}
	\includegraphics[page=2,width=\textwidth]{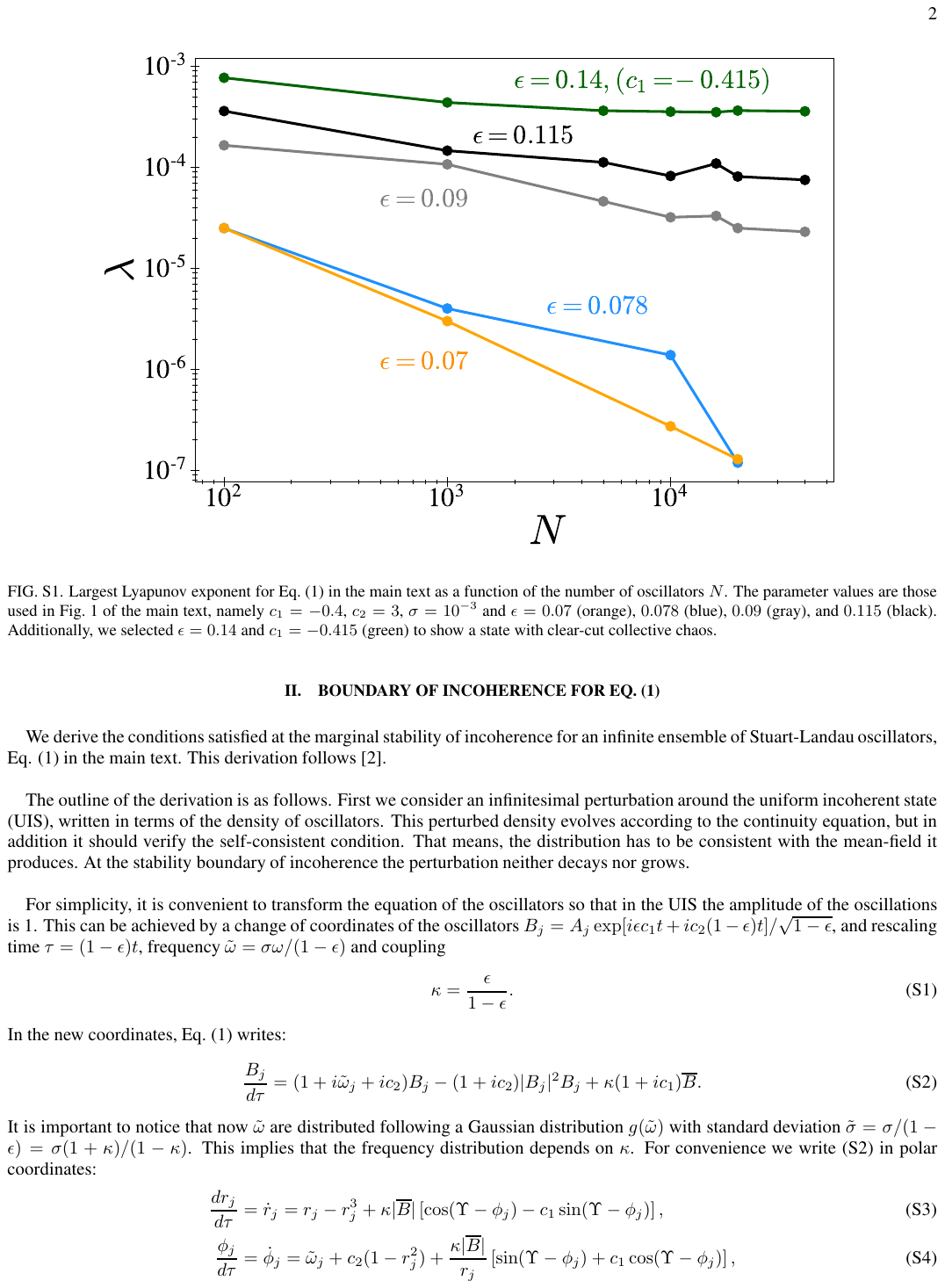}
\end{figure*}
\begin{figure*}
	\includegraphics[page=3,width=\textwidth]{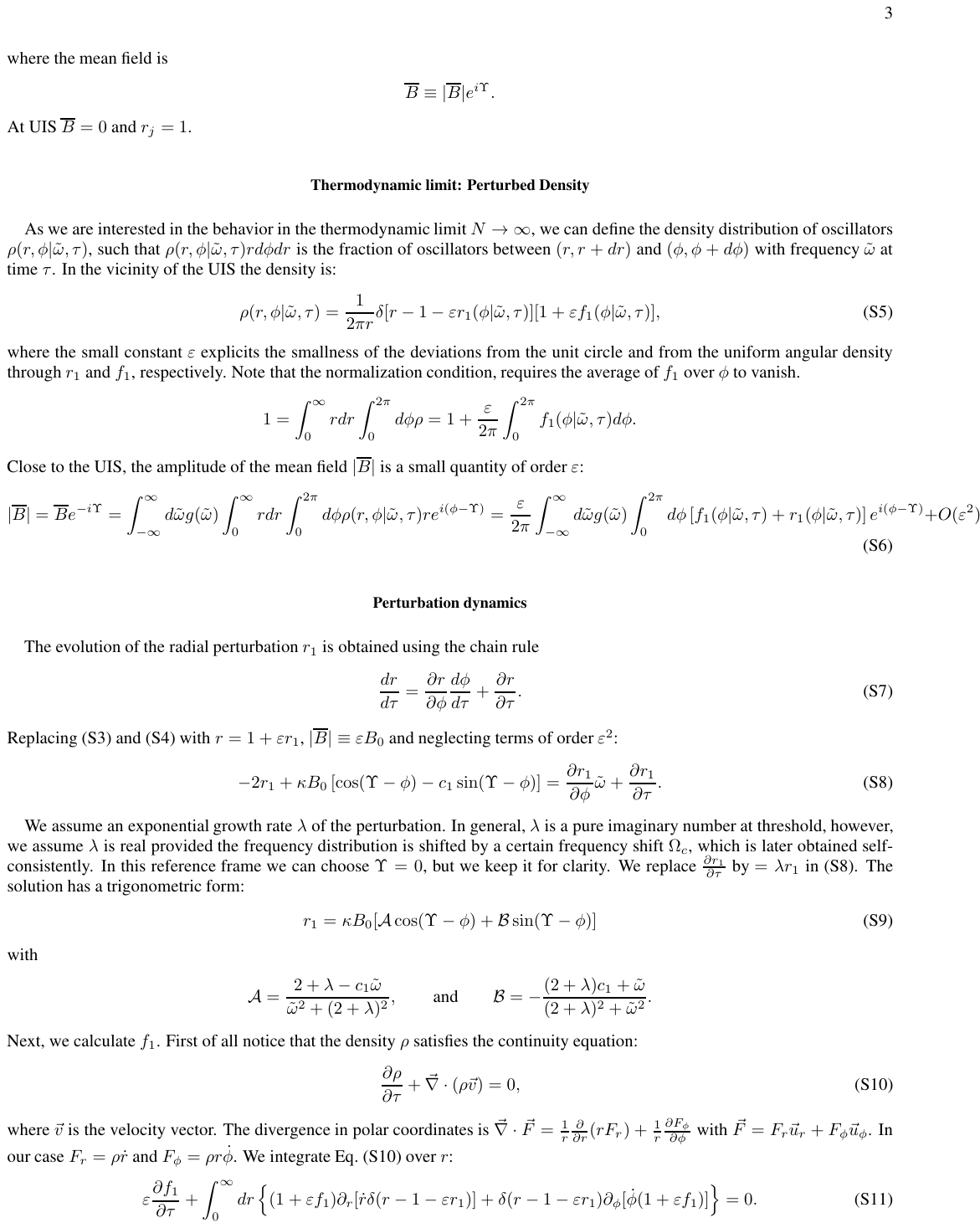}
\end{figure*}
\begin{figure*}
	\includegraphics[page=4,width=\textwidth]{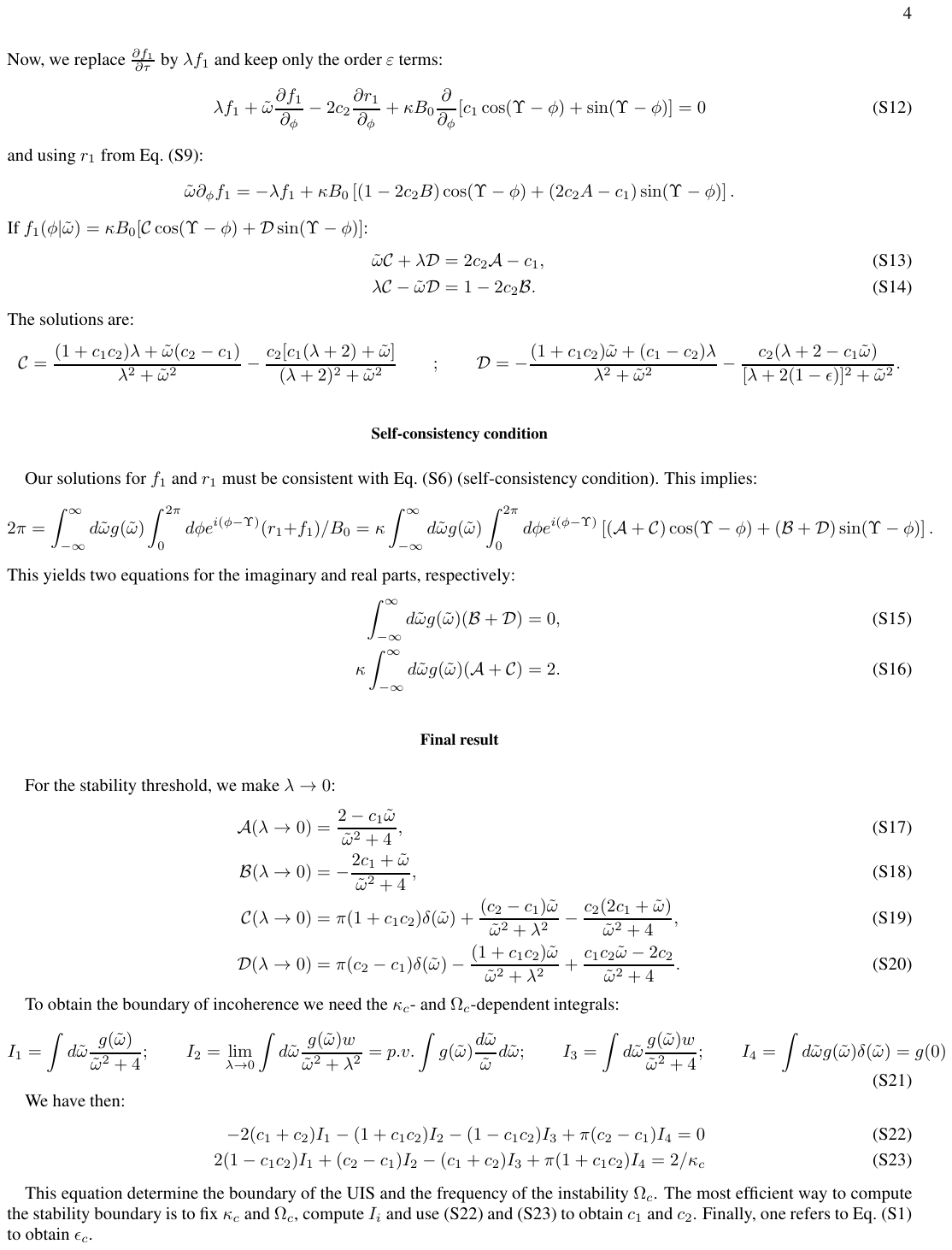}
\end{figure*}
\begin{figure*}
	\includegraphics[page=5,width=\textwidth]{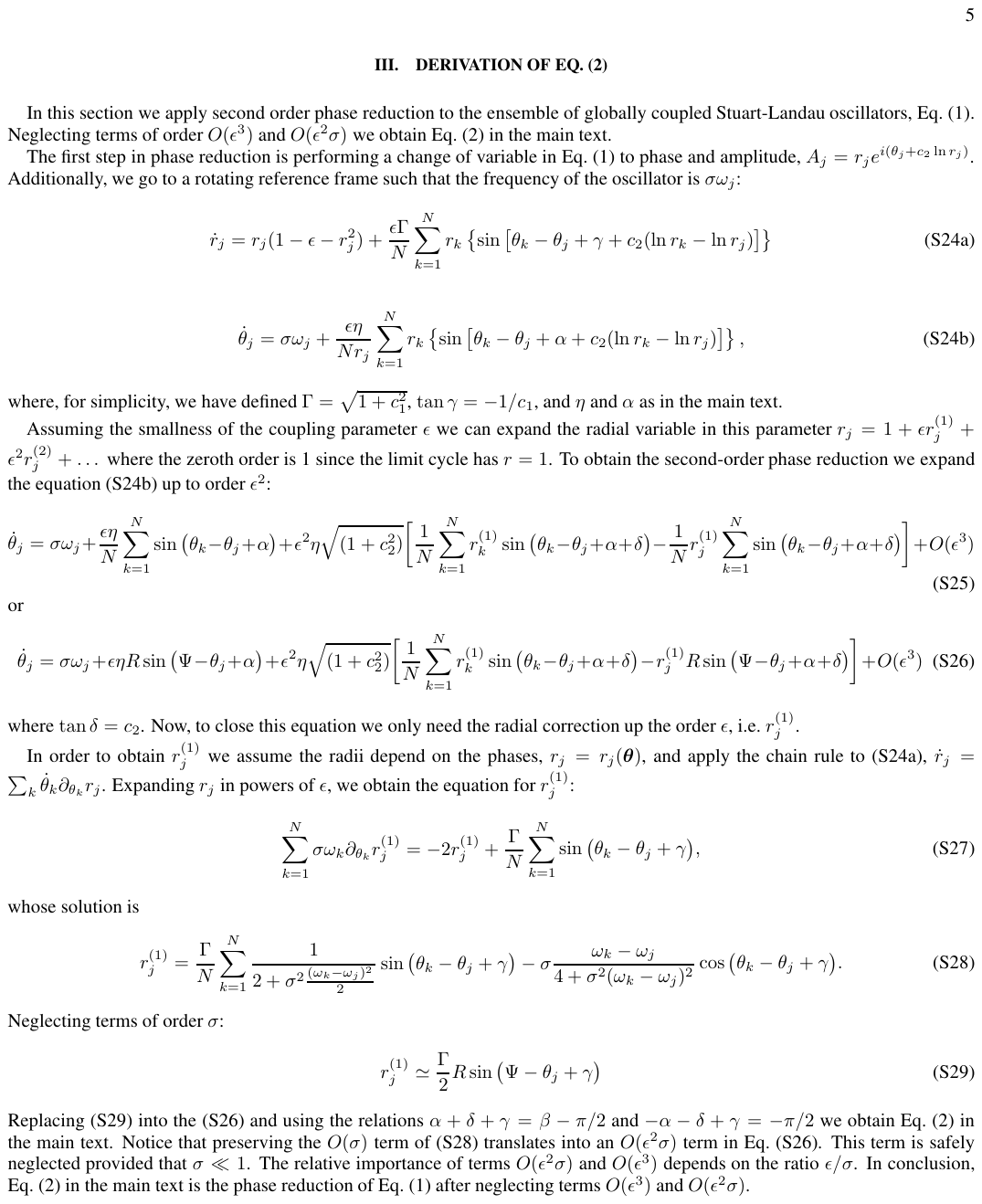}
\end{figure*}
\begin{figure*}
	\includegraphics[page=6,width=\textwidth]{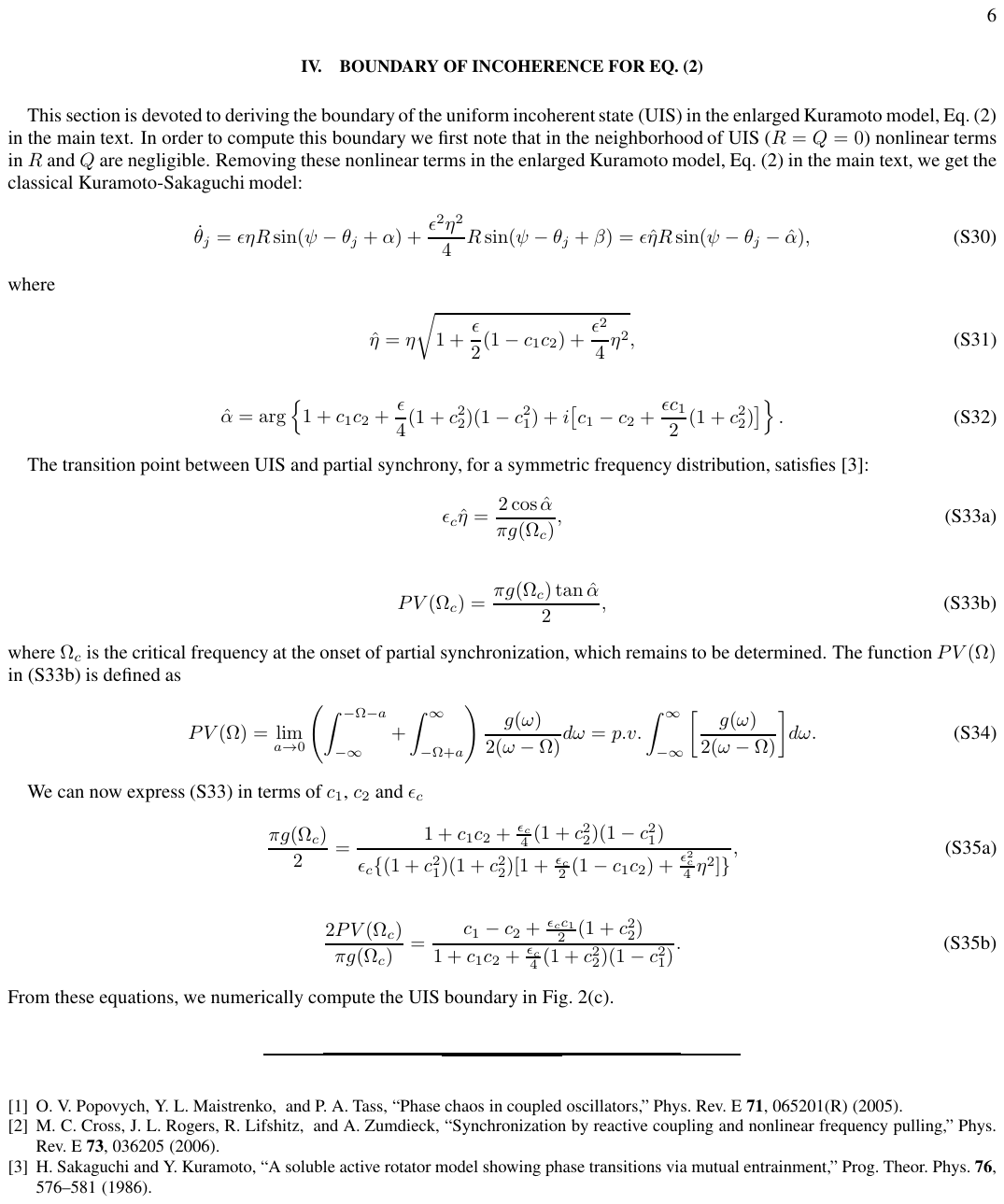}
\end{figure*}
\end{document}